\pdfoutput=1 
\documentclass[a4paper,english]{article}

\usepackage{graphicx}
\usepackage{color}
\usepackage{url,booktabs}
\usepackage{mdwlist}
\usepackage{subfig}
\usepackage{amsmath}
\usepackage{amssymb}
\usepackage[table]{xcolor}

\begin{document}

\title{Stegobot: building unobservable communication networks using social
  network behavior}

\author{Shishir Nagaraja\\
        Vijit Singh\\
        Pragya Agarwal\\
  Indraprastha Institute of Information Technology, New Delhi, India\\
  \tt{\{nagaraja, vijit, pragya\}@iiitd.ac.in}
  \and
  Amir Houmansadr\\
  Pratch Piyawongwisal\\
  Nikita Borisov\\
  University of Illinois at Urbana-Champaign, Urbana, IL, USA\\
  \tt{\{ahouman2,piyawon1,nikita\}@illinois.edu}
}

\date{}

\maketitle

\newcommand{\paragraphb}[1]{\vspace{0.03in}\noindent{\bf #1} }
\newcommand{\paragraphe}[1]{\vspace{0.03in}\noindent{\em #1} }
\newcommand{\paragraphbe}[1]{\vspace{0.03in}\noindent{\bf \em #1} }
\newcommand{\comment}{\textcolor{black}}
\newcommand{\nchange}{\textcolor{red}}
\newcommand{\ncomment}[1]{\textcolor{red} \itshape [#1]}
\newcommand{\ndelete}[1]{\textcolor{red}\sout{#1}}

\begin{abstract}
We propose the construction of an unobservable communications network using social networks. The communication endpoints are vertices on a social network. Probabilistically unobservable communication channels are built by leveraging image steganography and the social image sharing behavior of users. All communication takes place along the edges of a social network overlay connecting friends. We show that such a network can provide decent bandwidth even with a far from optimal routing mechanism such as restricted flooding. 

We show that such a network is indeed usable by constructing a botnet on top of it, called Stegobot. It is designed to spread via social malware attacks and
 steal information from its victims.  Unlike conventional botnets, Stegobot traffic does not introduce new communication endpoints between bots. We analyzed a real-world dataset of image sharing between members of an online social network. Analysis  of Stegobot's network throughput indicates that stealthy as it is, it 
 is also functionally powerful -- capable of channeling fair  quantities of sensitive data from its victims to the botmaster at ens of megabytes every month.
\end{abstract}

\section{Introduction}

Malware is an extremely serious threat to modern networks.  In recent
years, a new form of general-purpose malware known as {\it bots} has
arisen.  Bots are unique in that they collectively maintain
communication structures across nodes to resiliently distribute
commands and data through a {\it command and control} (C\&C) channel. The
ability to coordinate and upload new commands to bots gives the botnet
owner vast power when performing criminal activities, including the
ability to orchestrate surveillance attacks, perform DDoS extortion,
sending spam for pay, and phishing.


The evolution of botnets has primarily been driven by botnet responses
based on the principle of `whatever-works'.  Early botnets followed
a centralized architecture however the growing size of botnets led to
scalability problems.  Additionally, the development of mechanisms
that detect centralized command-and-control servers further
accelerated their
demise~\cite{binkley:sruti06,karasaridis:hotbots07,goebel:hotbots07}. This
led to the development of a second generation of decentralized
botnets.  Examples are Storm and
Conficker~\cite{stover:login07,stormpeacomm,porras:leet09} that are
significantly more scalable and robust to churn.



We believe that one of the main design challenges for future botnets
will be covertness --- the ability to evade discovery will be crucial
to a botnet's survival as organizations step up defense efforts. While
there are several covertness considerations involved, one of the most
important ones is hiding communication traces. This is the topic of
the present paper. We hope to initiate a study in the direction of
defenses against covert botnets by designing one in the first place.

We discuss the design of a decentralized botnet based on a model of
covert communication where the nodes of the network only communicate
along the edges of a social network. This is made possible by recent
advances in malware strategies. Social malware refers to the class of
malware that propagate through the social network of its victims by
hijacking social trust.  Instances include targeted surveillance
attacks on the Tibetan Movement~\cite{snoopingdragon} and the
non-targeted attack by the Koobface~\cite{koobface} worm on a number
of online social networks including Facebook~\cite{facebook}.

By adopting such a communication model, a malicious network such as a
botnet can make its traffic significantly more difficult to be
differentiated from legitimate traffic solely on the basis of
communication end-points. Additionally, to frustrate defense efforts
based on traffic flow classification, we explore the use of covert
channels based on information hiding techniques. What if criminals
used steganographic information hiding techniques that exploit human
social habits in designing botnets? Would it be possible to design
such a botnet?  Would it be weaker or stronger than existing botnets?
These are some of the questions we hope to answer in this paper.

The rest of this paper is organized as follows: in Section~\ref{sec:prelim}
we describe our threat model along with an overview on JPEG steganography primitives, which is essential in
the design of the social botnet introduced in this paper,
\textit{Stegobot}. In Section~\ref{sec:model} we describe the design
and construction of various components. We evaluate the use of of
image steganography in designing the command and control channel of
Stegobot using a real world dataset in Section~\ref{sec:stegoexpt};
and the routing mechanism in Section~\ref{sec:routing}. This is followed
by related work in Section~\ref{sec:related} and conclusions in
Section~\ref{sec:conclusions}.

\section{Preliminaries}\label{sec:prelim}

\subsection{Threat model}
We assume the threat model of a global passive adversary. Since a
botnet is a distributed network of compromised machines acting
cooperatively, it is fair to assume that the defenders will also
cooperate (ISPs and enterprises) and hence have a global view of
communication traffic (strong adversary).

We also assume that botnet infections are not detected. As with any
botnet Stegobot cannot withstand hundred-percent clean up of all
infected machines. However we expect it to easily withstand random
losses of a considerable numbers of bots. This assumption is due to
the fact that online social networks are often scale-free
graphs. In a seminal paper~\cite{AJB00}, Albert and
Barab{a'}si showed that scale-free graphs are highly robust to the
removal of randomly selected nodes. Indeed the real world social graph
considered in this paper (see dataset description in
section~\ref{sec:routing}) has a power-law degree-distribution with a
slope of $\gamma=2.3$.

\subsection{JPEG steganography}

%

A primary goal of this paper is to show that a botnet based on covert
channels can be constructed with a simple design and successfully
operated.  We use JPEG steganography to construct communication
channels between the bots. We now review the main results in JPEG
steganography that are of relevance to this paper. A full discussion
on the relative merits and demerits of various design choices is
defered until section~\ref{sec:related}.

We considered the JSteg scheme \cite{jsteg,provos:sp03} but the
resulting steganographic capacity of the communication is rather low;
steganographic images are detectable~\cite{Lee:IH07} even at low
embedding rates of $0.05$ bits per non-zero non-one coefficients. A
better scheme is proposed by Fridrich et al.~\cite{fridrich:mmsec07}
who showed that the average steganographic capacity of grayscale JPEG
images with quality factor of $70$ can be approximated to be $0.05$
bits per non-zero AC DCT coefficient. The most recent scheme based on
the same principle (of minimal distortion embedding) as the Fridrich
scheme is the YASS~\cite{Solanki:IH07} scheme, which has been shown
undetectable at payloads of $0.05$ bits per non-zero DCT coefficient.

%

\section{Stegobot construction} \label{sec:model}
A botnet is a distributed network of a number of infected
computers. It is owned by a human controller called the {\bf
  botherder} and consists of three essential components: the
botmaster(s), the bots, and the Command and Control (C\&C)
channel. {\bf Bots} are compromised machines running a piece of
software that implement commands received from one or more {\bf
  botmasters}; they also send {\bf botcargo} -- information acquired
by the bot such as the result of executing botherder commands -- to
the botmaster. Botmasters refer to compromised machines that the
botherder interacts with in order to send commands via a {\bf C\&C}
channel. The botmaster sends instructions to the bots to carry out
tasks and receives botcargo sent back to it by the bots.

\subsection{Design goals}
A distinguishing feature of Stegobot is the design of the
communication channel between the bots and the botmaster. Stegobot is
designed for {\it stealth}, therefore we do not wish to include {\it
 any} C\&C communication links between computers that do not already
communicate.

A further goal is to design {\it probabilistically unobservable}
communication channels connecting the botmaster and the bots. If the
C\&C communication is unobservable then botnet detection can be
significantly more difficult than where communication is not
hidden. This is because in the latter case, traffic-flow signatures
and the changes in the structure of traffic connectivity induced by
the presence of the botnet can lead to easier detection and removal of
the botnet~\cite{gu:usenix-security08,botgrep}.

\begin{figure}[!t]
\begin{centering}
\includegraphics[scale = 0.95]{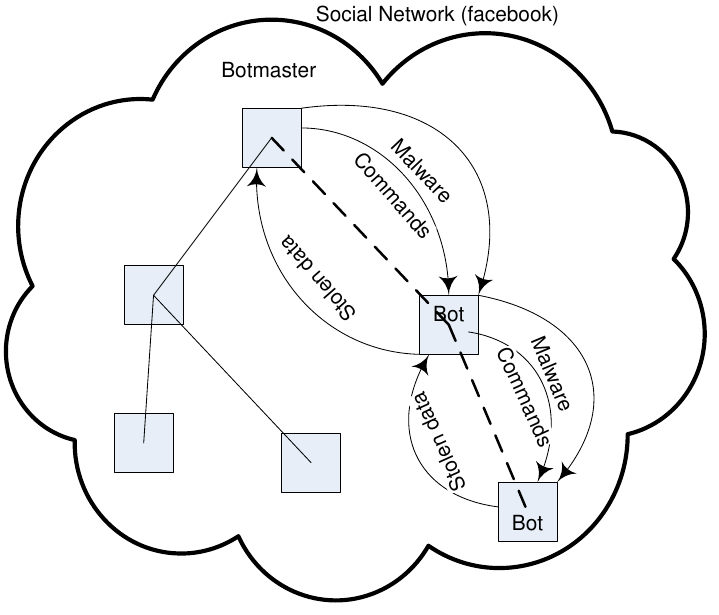}
\caption{The topology of the Stegobot botnet}
\label{fig:botnet_topology}
\end{centering}
\end{figure}

\subsection{Malware propagation and bots}
The first step in botnet creation is malware deployment. The malware
is an executable which infects the machine and performs the activities
necessary of a bot. Stegobot is designed to infect users connected to
each other via {\it social links} such as an email communication
network or an online social network that allows friends to exchange
emails. The propagation of malware binaries takes place via
social-malware attacks~\cite{snoopingdragon}.

Social-malware attacks refer to the use of carefully written email
lures to deliver botnet malware combined with the use of email
communication networks to propagate malware. This works when the
attackers take the trouble to write emails that appear to come from
the co-workers or friends of the victim (social phish). A more
effective attack is to replay a stolen email containing an attachment
that was genuinely composed by a friend back to the victim after
embedding a malicious payload within the attachment.

Once the attacker secures an initial foothold (deploy the malware on
at least one victim's machine), the attacker can expand the list of
compromised machines with little additional effort. Whenever one of
the initial set of victims sends an email containing an attachment to
one of their colleagues, the bot quickly embeds a malicious payload to
the attachment. Upon opening the attachment, the receiver's computer
also gets infected and the process continues. Therefore once a single
user is compromised (and the compromised machine continues to be
operated for sending emails), the attacker can recruit additional bots
in an automated fashion.  Indeed this was the modus operandi behind
the Ghostnet surveillance attacks on both Google and the Tibetan
administration in 2009~\cite{snoopingdragon}.

Of course the attacker's attempts at composing email lures can fail
with non-zero probability. However this exercise needs to succeed only
once (as explained in the previous paragraph) to generate a botnet
containing thousands of nodes, and the risk of failure is offset by
targeting multiple people within a social group.

\subsection{Bots}
In Stegobot, bots possess a pre-programmed list of activities such as
harvesting email addresses and passwords, or credit card numbers or
simply to log all keystrokes. Alternatively, in a more flexible design
the bots execute commands received from the botmaster. For instance,
bots receive search keywords from the botmaster and respond with
matches from the filesystem, as in the case of the Tibetan
attacks~\cite{snoopingdragon}.

As explained in the previous paragraph, Stegobot spreads along the
social network of its victims.  While we have used emails to explain
social-malware attacks, the attacks are by no means restricted to
email communication networks alone; online social networks are equally
good targets. For instance, Koobface~\cite{koobface} is a worm that
propagates on Facebook over social links, demonstrating that migrating
from conventional email to social network messaging does not insulate
users from social malware attacks. Further, it is noteworthy that
Facebook is adding email extensions to its existing service; and
Google added a social networking service --- Google Buzz --- to its
popular email service in 2010. This allows bots to communicate with
each other and the botmaster over the social network as explained in
the next section.

\subsection{Message types} \label{sec:msgtypes}
Stegobot uses two types of message constructions. First, {\bf Bot-commands}
are broadcast messages from the botmaster. Examples
include search strings for file contents or within keylogged data.

Second, \textit{botcargo} messages return information requested by the
botmaster such as files matching search strings. Botcargo messages can
be divided further into two types: locally generated (\textit{botcargo-local}) or forwarded messages (\textit{botcargo-fwd}) on a
multi-hop route to the botmaster.

\subsection{Communication channel} \label{sec:cc}

In Stegobot, we use the images shared by the social network users as a
media for building up the C\&C channel. Specifically, we use image
steganography techniques to set up a communication channel within the
social network, and use it as the botnet's C\&C channel.

A bot running on a computer can communicate with a bot running on a
different computer if both the computers are being used by people
connected by an edge in the social network. The social network acts as
a peer-to-peer overlay over which the information is transferred from
each bot to the botmaster. In Stegobot, information between bots must
only be transferred using steganographic channels. In our case, this
channel is constructed by hiding the botcargo within images using
standard techniques reviewed in earlier sections. By keeping the size
of the botcargo within certain limits, it is possible to make the
presence of bot communication difficult to discover by examining the
communication channel alone (section~\ref{sec:stegoexpt}).

\begin{figure*}[!t]
\begin{centering}
\includegraphics[width = 4.8 in]{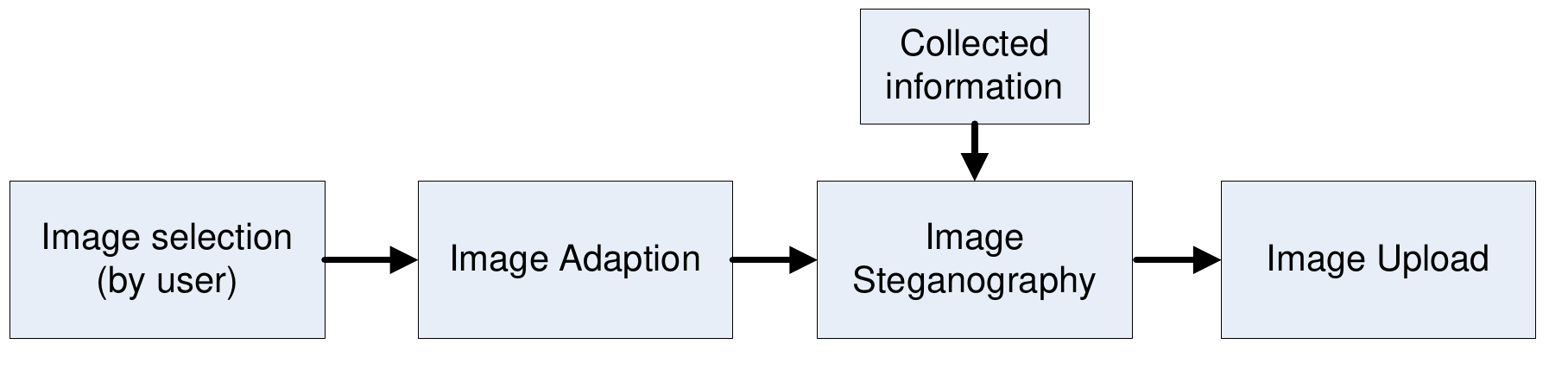}
\caption{Process of sending a one-hop message}
\label{fig:bot_insertion}
\end{centering}
\end{figure*}

\paragraph{One-hop communication} takes place according to a {\it push-pull}
model.  Consider the example of Facebook (see
figure~\ref{fig:bot_insertion}). When a user {\it pushes} (uploads) an
image to Facebook from an infected host, the bot intercepts the image
and inserts the botcargo into the image using an image steganography
technique as previously discussed. In our prototype this was done by
uploading botcargo into all pictures on the victim's computer; a more
practical approach might be to concentrate on a subset of directories
where the user stores pictures. Upon completion of image upload, all
the neighbors of the user are notified (by Facebook). When a neighbor
of the publisher logs into Facebook from an infected machine and views
the picture, the bot {\it pulls} (intercepts) the image and extracts
the steganographically embedded botcargo from the image. All botcargo
is finally destined for the botmaster who downloads the cargo by
viewing newly posted pictures from her neighbors. When the botmaster
intends to issue a command, she does so by preparing a botcargo
message and uploading it to her Facebook account. It is worth noting
that Facebook presently downloads all the images on to your computer
automatically when a Facebook page is visited; the embedded images
don't need to be clicked on by the victim for botcargo transfer.

While the communication channel used in our design and experiments is
based on Facebook, any social communication mechanism involving rich
content can be utilized in its place. In theory, blocking access to
Online Social Networks (OSNs) will stop Stegobot. In practice, efforts
to limit access is not easy since the use of OSNs for furthering
business goals is on the increase. Additionally, such measures are easily
circumvented by determined users leveraging open anonymizing proxies.

\paragraph{Multi-hop communication:} In Stegobot, routing is based on a
 very simple algorithm namely {\bf restricted flooding}.

\paragraph{Congestion control:} Each bot maintains a bandwidth
\textit{throttle} which is used to control the ratio of
\textit{botcargo-local} to \textit{botcargo-fwd} messages.

\paragraph{Metrics:} We measure the effectiveness of the routing strategy using a set of metrics.
\begin{itemize}

\item {\it Channel efficiency} the percentage of \textit{botcargo-fwd}
  messages that reach the botmaster averaged over all bots.

\item {\it Channel bandwidth} is similar to efficiency, but it is the
  absolute number of \textit{botcargo-fwd} messages that reach the
  botmaster averaged over all bots.

\item {\it Duplication count} is the number of duplicate \textit{botcargo-fwd}
  messages received by the botmaster.

\item {\it Botnet bandwidth} is the total number of
  \textit{botcargo-fwd} reaching the botmaster every month excluding
  duplicates.

\end{itemize}

\section{Experiments}
In order to convince ourselves that a Stegobot deployment could indeed
be profitably operated in a real world setting, we performed a number
of experiments which are detailed below.

\subsection{Steganography experiments}
\label{sec:stegoexpt}
We use YASS \cite{Solanki:IH07} as the image steganography scheme of
the C\&C channel over the Facebook social network. Facebook's image
processing can interfere with the bots' steganographic communication
channel. In order to minimize this, the bot performs an image adaption
process as follows before embedding a payload: 1) each image is
converted to the JPEG format, 2) images are resized to meet the
maximum resolution limits performed by Facebook, i.e.,
$720\times720$ \footnote{More recently, Facebook is allowing uploading
  of higher-resolution images that increase the steganographic
  capacity at least 10 times based on our preliminary
  experiments}. This is performed keeping the aspect ratio of the
images.

We use a database of 116 different images to perform our
experiments. In each experiment an image is adapted to Facebook
constraints, as mentioned before, and then the hidden information is
embedded into that image using YASS scheme. The stego image is then
uploaded into Facebook through a Facebook user account, and then
downloaded from the Facebook using another Facebook account. Finally,
the downloaded image is evaluated by the YASS detector described in
\cite{Solanki:IH07} in order to extract the hidden message. To
evaluate the robustness of our steganographic process we calculate the
bit error rate (BER) metric which is defined as the ratio of error
message bits to the total number of message bits for each image.

Table~\ref{tab:steg1} summarizes the average of the BER parameter
(over all of the images) for different metrics of YASS scheme.  $Q$ is
the quality factor of YASS scheme and represents the amount of
compression performed by YASS during the steganography process. $Q$
has a range of $[0,100]$ and directly impacts the quality of the stego
image, i.e., higher $Q$ results in images with higher quality/size.
Based on the results of our experiments, Facebook's uploading process
is equivalent to the application JPEG compression over the image with
a quality factor of $Q_f$.  For $Q> Q_f$ Facebook applies extra
compression on the image which results in loosing some of hidden
information bits. On the other hand decreasing $Q$ results in lower
number of bits being inserted by the YASS scheme. So, there should be
an optimum value for $Q$ within the range of $[0,100]$ which minimizes
the BER rate, i.e., maximizes the robustness to Facebook
perturbations. As table~\ref{tab:steg1} shows the BER values are
minimized for a $Q=75$, hence we approximate the quality factor of the
Facebook compression to be $Q_f \approx 75$.

We also investigate the effect of the redundancy parameter of YASS,
$q$, on the BER. The parameter $q$ represents the number of times an
information bit is repeated inside an image by the YASS
scheme. Intuitively, we expect that larger $q$ results in reducing the
BER, since more redundant bits can help better in reconstructing a
noisy message; this is confirmed through our experiments as
table~\ref{tab:steg1} shows. In fact, the $q$ parameter makes a
tradeoff between robustness and steganographic capacity: increasing
$q$ improves robustness by reducing BER while it also reduces the
number of data bits inserted by the YASS
scheme. Table~\ref{tab:databits} shows the number of bits inserted by
YASS for different values of $q$.

Our experiments show that a small number of image, namely \textit{bad
  images}, are responsible for a majority of errors in the average
BER. Excluding these images in the steganography process can
significantly reduce average BER. We define and use a metric,
\textit{SelfCorr}, in order to decide whether an image is  'bad' or
'good'. The \textit{SelfCorr} metric evaluates the cross
correlation of an image by a noise-filtered version of itself. We
declare images with \textit{SelfCorr}$>0.9964$ as 'bad'
images. Table~\ref{tab:steg2} illustrates the BER results after
excluding the small number of 'bad' images determined by the
\textit{SelfCorr} metric. As can be seen, the average BER is
significantly improved, e.g, the average BER is 0 for $q \geq 12$.

\begin{table}[ht]
\centering
\caption{Average BER (over 116 images) without removing 'bad images'}
\label{tab:steg1}
\begin{tabular}{|c|c|c|c|c|c|c|c|c|c|c|}
  \hline
q	& 2&	4&	6&	8&	10&	12&	14&	16&	18&	20\\ \hline
Q=65&	0.3073&	0.1320&	0.0520&	0.0227&	0.0097&	0.0047&	0.0022&	0.0010&	0.0006&	0.0003\\
Q=70&	0.2966&	0.1318&	0.0529&	0.0219&	0.0096&	0.0049&	0.0025&	0.0010&	0.0005&	0.0002\\
\rowcolor{gray} Q=75&	0.3015&	0.1557&	0.0680&	0.0283&	0.0101&	0.0067&	0.0027&	0.0010&	0.0004&	0.0000\\
Q=80&	0.3086&	0.1839&	0.0846&	0.0347&	0.0143&	0.0089&	0.0034&	0.0015&	0.0008&	0.0000\\
Q=85&	0.3512&	0.2618&	0.1777&	0.0854&	0.0372&	0.0183&	0.0127&	0.0053&	0.0024&	0.0013\\
Q=90&	0.4287&	0.3917&	0.3639&	0.3390&	0.3146&	0.2906&	0.2567&	0.2122&	0.1591&	0.1262\\
\hline
\end{tabular}
\end{table}

\begin{table}[ht]
\centering
\caption{Number of bits inserted in each image for different values of $q$}\label{tab:databits}
\begin{tabular}{|c|c|c|c|c|c|c|c|c|c|c|}
  \hline
   q	& 2&	4&	6&	8&	10&	12&	14&	16&	18&	20\\ \hline
  Data bits & 40280&	20140&	13426&	10070&	8056&	6173&	5754&	5035&	4475&	4028\\
\hline
\end{tabular}
\end{table}

\begin{table}[ht]
\centering
\caption{Average BER after removing 'bad images'}\label{tab:steg2}
\begin{tabular}{|c|c|c|c|c|c|c|c|c|c|c|}
  \hline
  q	& 2&	4&	6&	8&	10&	12&	14&	16&	18&	20\\ \hline
Q=65&	0.2945&	0.1088&	0.0311&	0.0092&	0.0022&	0.0002&	0.0000&	0.0000&	0.0000&	0.0000\\
Q=70&	0.2836&	0.1105&	0.0340&	0.0095&	0.0016&	0.0002&	0.0000&	0.0000&	0.0000&	0.0000\\
\rowcolor{gray} Q=75&	0.2892&	0.1372&	0.0492&	0.0136&	0.0011&	0.0001&	0.0000&	0.0000&	0.0000&	0.0000\\
Q=80&	0.2977&	0.1686&	0.0662&	0.0175&	0.0020&	0.0003&	0.0000&	0.0000&	0.0000&	0.0000\\
Q=85&	0.3436&	0.2512&	0.1631&	0.0646&	0.0165&	0.0029&	0.0012&	0.0000&	0.0000&	0.0000\\
Q=90&	0.4255&	0.3877&	0.3589&	0.3331&	0.3074&	0.2823&	0.2464&	0.1978&	0.1396&	0.1035\\
  \hline
\end{tabular}
\end{table}



\subsection{Routing results} \label{sec:routing}

Combining social-malware with steganographic channels yields a covert
botnet where new bots are recruited as infections spread along the
edges of the social network, while existing bots communicate using the
well understood image based steganographic channels. In this section,
we study the routing capabilities of such a botnet using a real-world
example.

\paragraph{Dataset:} We chose to study the Flickr\footnote{Unfortunately, we did
  not have access to the Facebook topology or the upload patterns of
  users.} social network~\cite{flickr}, an online friendship network
that facilitates image sharing. We crawled the Flickr website and
downloaded on a fraction of the Flickr social network. Specifically,
our dataset contains 7200 nodes (people), the social network edges
(online friendship relations) between them, and the number of images
posted per person per month. The dataset corresponds to user activity
on Flickr over a period of $40$ months. The Flickr dataset will be made
available on our website for the research community.

In our simulation, each bot node generates $K$ \textit{botcargo-local}
(see section~\ref{sec:msgtypes}) messages per month. $K=10$
corresponds to say ten files that the bot plans to route to the
botmaster across the social overlay network. $ttl$ is fixed at
$log(N=7000) \approxeq 3$ hops. Each bot reserves a minimum of $5\%$
of node bandwidth to forward \textit{botcargo-fwd} messages received from
neighbors. Further, we assume \textit{bot-command} messages broadcast
from the botmaster at a rate of one message per month. This means that
the botmaster can instruct her bots to change operation no more than
once a month.

Stegobot's infection strategy is based on social malware attacks. In
our experiments, we have assumed an infection rate of 50\%. While this
number might appear high to some readers, it is actually a
conservative estimate; social-malware has been known to have infection
rates approaching 90-95\% in real-world attacks~\cite{snoopingdragon}.

\paragraph{Botcargo preparation:} Each bot gathers botcargo (both from
the host as well as from its neighbors). It then encodes as much of
the botcargo in a single image as allowable according to a detection
threshold $\ell$ bits. The practically possible values for the number
of bits is given in table~\ref{tab:databits} and a discussion in
section~\ref{sec:stegoexpt}.

%
%
\paragraph{Routing:} In Stegobot, routing is carried out by restricted
flooding. Each bot publishes (floods) botcargo to all neighbors
(joined the botnet) within $ttl$ hops in the social network. Finally,
the botmaster receives botcargo through the one of its infected
neighbors. We assume that the botmaster is a randomly chosen node in
the network. For each of the graphs below, we averaged the results
over fifty different botmaster nodes.

\begin{figure*}[!t]
\begin{centering}
\includegraphics[scale=0.25]{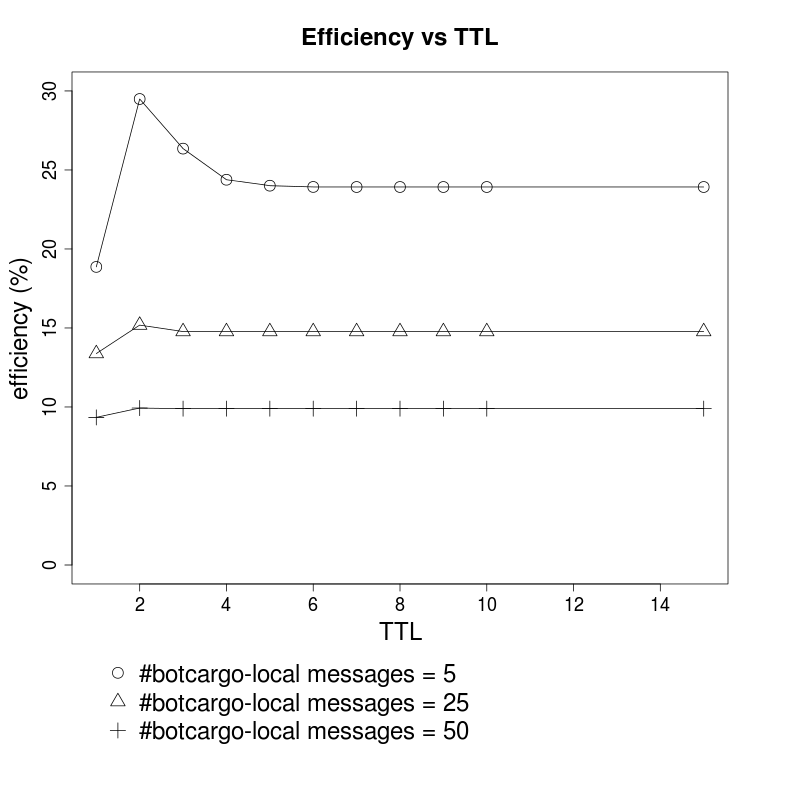}
\caption{Average channel efficiency against ttl}
\label{fig:ttl}
\end{centering}
\end{figure*}

Figure~\ref{fig:ttl} shows the efficiency of botcargo transmission for
increasing amounts of $ttl$ and various numbers of \textit{botcargo-local} messages. For $K=5$ \textit{botcargo-local} messages,
the efficiency peaks at $30\%$ and decreases and then stabilizes for
higher $ttl$ values as the resulting increase in the number of \textit{botcargo-fwd} messages begins to cause congestion. Congestion
effects are also felt when the number of \textit{botcargo-local} messages
increase even at a smaller $ttl$. This justifies our intuition for
using $ttl=log(N)$ where $N$ is the number of infected nodes in the
botnet.

In restricted flooding, high-degree nodes in the topology
play the role of hubs and are able to pull and collect large amounts
of botcargo. As such they become a natural point where stolen
information is collected and can then be siphoned off to the
botmaster.

\begin{figure*}[!t]
\begin{centering}
\subfloat[Normalized Bandwidth and Efficiency]
         {\includegraphics[width=0.49\textwidth]{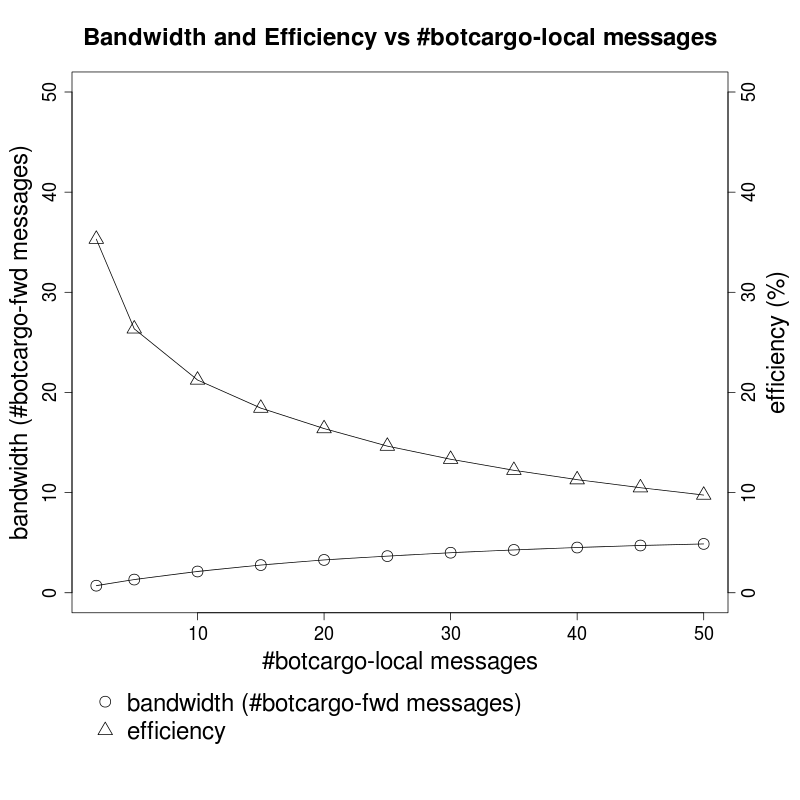}}
         \hfill
\subfloat[Duplication]
         {\includegraphics[width=0.49\textwidth]{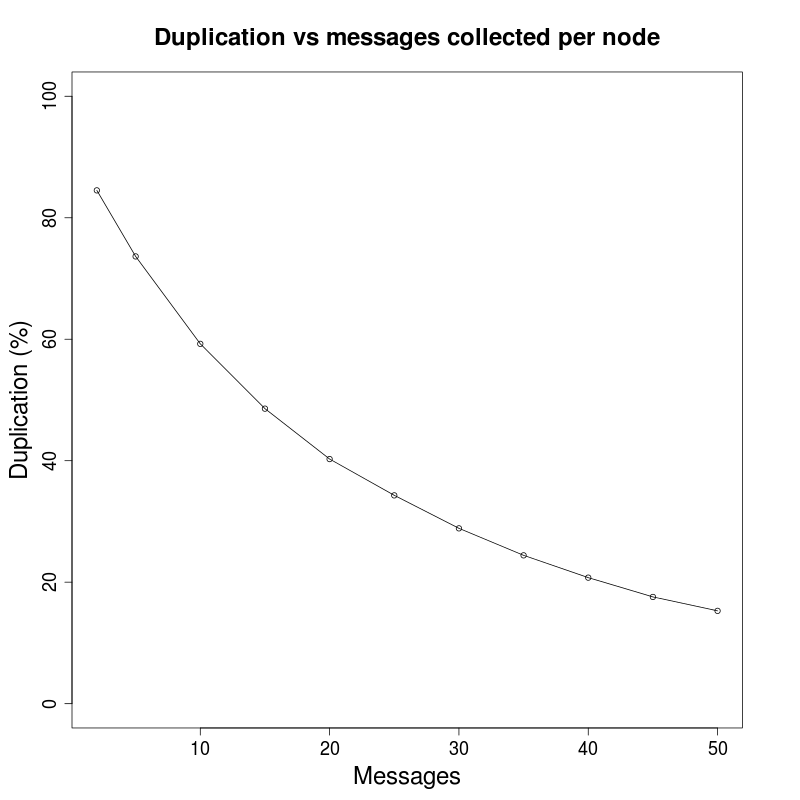}}
\caption{Communication channel bandwidth and efficiency}
\label{fig:bweff}
\end{centering}
\end{figure*}

\paragraph{Channel Bandwidth and Efficiency:} Figure~\ref{fig:bweff} shows
 the bandwidth and efficiency of the communication channel in the
 average case. Figure~\ref{fig:bweff}.a shows the monthly average
 number of \textit{botcargo-fwd} messages received by the botmaster
 (normalized by the size of the botnet) for various amounts of
 \textit{botcargo-local} messages collected per bot (constant across
 bots). Figure~\ref{fig:bweff}.a also shows the average efficiency of
 the communication channel from a bot to the botmaster as the size of
 the botcargo changes. The network seems to operate at an average
 efficiency of 30\% of collected botcargo reaching the botmaster when
 $K=2$ (\#botcargo per bot per month). This decreases with increase in
 $K$ although the absolute number of messages delivered at the
 botmaster increases marginally from $.75$ per bot for $K=2$ to $2.5$
 per bot for $K=10$. Further increases result in even more marginal
 increases as the effects of congestion result in decreasing routing
 efficiency. A positive effect of increasing per node botcargo
 collection sizes ($K$) is the reduction in duplicate messages
 reaching the botmaster. This is shown in figure~\ref{fig:bweff}.b,
 the proportion of duplicate messages rapidly decreases until $K=10$
 and further reduces to $40\%$ at $K=20$. We observe that the positive
 effects of duplication reduction correspond with an increase in
 normalized bandwidth as the number of \textit{botcargo-local}
 messages collected per node increase.

\begin{figure*}[!t]
\begin{centering}
\subfloat[Botcargo delivered]
         {\includegraphics[width=0.49\textwidth]{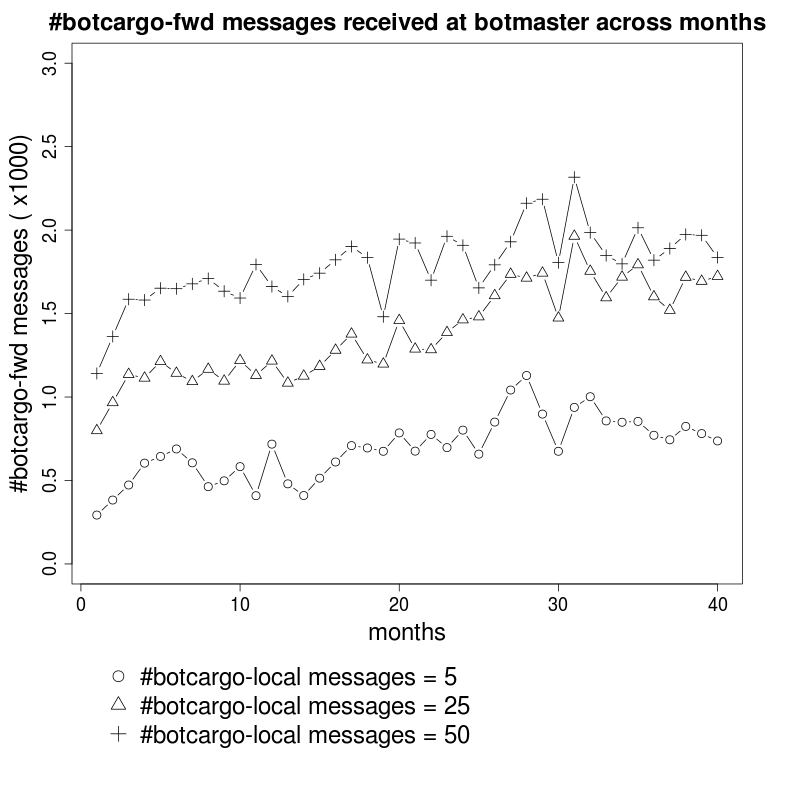}}
         \hfill
\subfloat[Cumulative botcargo delivered]
         {\includegraphics[width=0.49\textwidth]{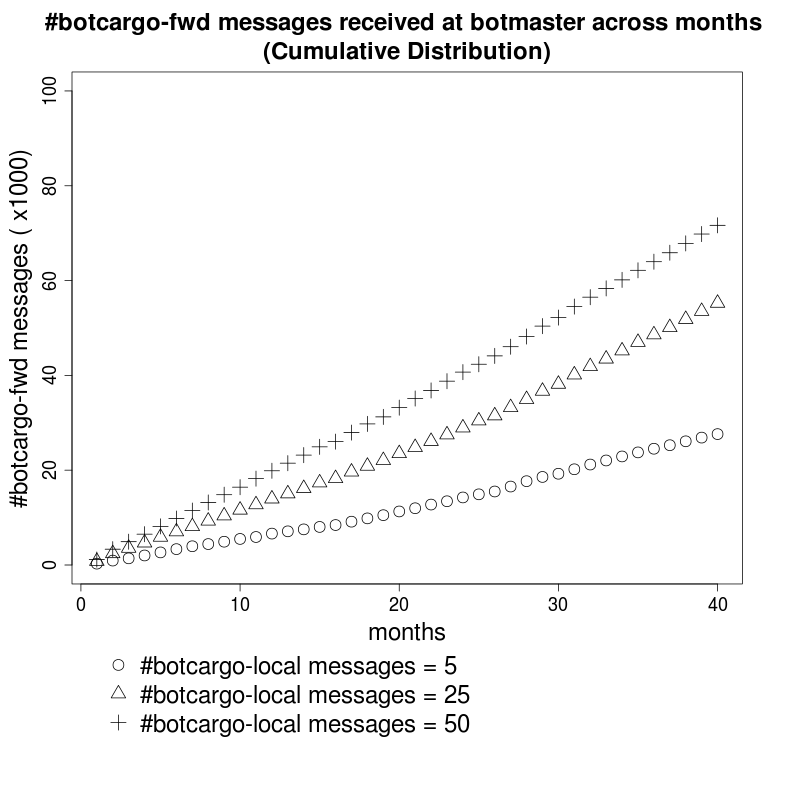}}
\label{ttl}
\caption{Experimental results for the number of delivered batcargo}
\label{fig:botcargo}
\end{centering}
\end{figure*}

The main result of our experiments is shown in
figure~\ref{fig:botcargo}. Figure~\ref{fig:botcargo}.a shows the
average number of botcargo messages delivered to the botmaster. This
shows an increasing trend. This can be traced to the increasing number
of users and the number of average number of photo updates per user
increase over the months in our dataset. The sharp drops and increases
are related to routing performance under {\bf churn}, when a few large
uploaders suddenly stop using uploading for certain periods of time,
or dormant users being uploading in larger numbers (say from one-two
images to twelve-fifteen images per
month). Figure~\ref{fig:botcargo}.b indicates the cumulative amount of
traffic received by the botmaster over the years and gives a sense of
the total amount of sensitive material she can steal and the long-term
trends. Combining the total number of messages reaching the botmaster
($18000$ \textit{botcargo-fwd}) with the number of bits embedded in each
message, we obtain a monthly bandwidth of between 21.60MB/month in the
average case ($q=8$) to 86.13MB ($q=2$) for lower interference from
the image adaption process.

%

Overall, it is easy to see that even with a simple and naive routing
algorithm such as restrictive flooding, the botmaster is easily able
to collect around $10\%$ of the total amount of stolen
information. With a slightly more sophisticated algorithm that
exploits the presence of medium and high degree hub nodes as
super-peers, one could design a better routing algorithm. For
instance, in the current implementation all nodes behave the same way,
hence hub nodes also locally flood all the botcargo they receive. This
is replayed back and forth between hubs and the rest of the network
causing severe congestion. By ensuring that super-peers carefully
route incoming botcargo only to other super-peers, we believe it
should be possible to significantly improve network throughput.

\section{Related work}\label{sec:related}


Most current botnets use a peer-to-peer
architecture~\cite{stormpeacomm,porras:leet09} which improves
robustness and scalability. Botnet detection techniques exploit
inter-bot interaction patterns~\cite{botgrep} or exploit the
statistical characteristics of traffic
flows~\cite{gu:usenix-security08,yen:dimva08} to localize bots. Both
these approaches require access communication traffic between the
bots. By using (probabilistically) unobservable communication
channels, Stegobot evades all these detection approaches.
%

The work closest to ours is the work of Nappa et
al.~\cite{skype-botnet} who describe the design of a resilient botnet
using the Skype protocol for inter-bot communication. The use of Skype
for VoIP communication is popular and is hence difficult to block
without annoying legitimate users. By hijacking active (logged in)
Skype sessions, the botnet is able to bypass firewalls that might
otherwise prevent bots from directly communicating with each
other. Our design goes a lot further due to the unobervability
properties of our communication channel. Unlike the design of Nappa et
al., we do not add new connection end-points -- no communication
between user-accounts (bots) that do not already communicate, and no
additional communication is introduced beyond what that users already
exchange, resulting in a stealthy design.

\subsection{JPEG steganography}
Practical steganography schemes are based either on heuristic methods
or provide some provable security based on some specific model. One of
the first practical steganography schemes for the JPEG images is the
JSteg scheme \cite{jsteg,provos:sp03}. It is based on using the Least
Significant Bit (LSB) components of the quantized DCT
coefficients. More specifically, the message bits are simply replaced
with the LSBs of the DCT coefficients of an image, considering some
exclusions for preserving the image quality. The embedding path for
the LSBs was originally sequential while the use of pseudo-random path
was suggested in later implementations. Even with pseudo-random path
the LSB steganography techniques are shown to be detectable through
different kind of
attacks~\cite{Westfeld:IH00,Yu04:ICPR04,Lee:IWDW06,Lee:IH07} that
exploit artifacts made in the first order statistics of the DCT
coefficients.


These attacks led the next generation of the JPEG steganography
schemes, namely statistical restoration-based schemes, to consider
preserving statistical behavior of the cover images
\cite{Solanki:ICIP06}. The main idea is to divide the cover image into
two disjoint parts, which one part is used to embed the message and
the other part is used to make corrections in order to preserve the
selected statistical behavior of the image. A related approach for
preserving the statistical behavior is used in the Model Based
Steganography~\cite{Sallee:IWDW03}, where some specific \textit{model}
is preserved for the DCT coefficients.

As an example of the heuristic steganography schemes we can mention
the F5 scheme \cite{Westfeld:IH01}, which was developed to address the
detectability of the LSB-based embedding schemes. By decreasing the
absolute value of the coefficients by $1$ and using some other tricks
the F5 scheme avoids the obvious artifacts on different features of
the image. To increase the embedding efficiency F5 uses a coding
scheme, namely Matrix Embedding.

Another approach for steganography, recently attracting more
attention, is the minimal distortion
embedding~\cite{Fridrich:GS05,Kim:IH06}. These schemes focus on
increasing the embedding efficiency by decreasing the embedding
distortion.  Newman et al.\ in~\cite{Newman:IH02} propose
JPEG-compatibility-steganalysis resistant method, which embeds the
message into the spatial domain of the image before performing the
JPEG compression. YASS~\cite{Solanki:IH07} is a more recent
scheme based on the approach of minimal distortion embedding.
%
%
%

\section{Conclusions}\label{sec:conclusions}
The essence of communication  security lies not merely in
protecting content but also unobservability.  In this paper, we have
presented and analyzed the design of a covert botnet using
unobservable communication channels that aims to steal sensitive
information. The proposed botnet deploys innovative social malware
infection strategies to create an overlay network over the social
communication network of victims. A critical aspect of our design is
the use of image based steganographic techniques to hide bot
communication and make it indistinguishable from image noise. While
techniques for image steganography are well known, we go one step
further to show that it is possible to design an effective covert
network by exploiting the social network connecting users and the
social habits of individual users.

%

\subsection*{Acknowledgements}
The authors would like to thank Anindya Sarkar for providing the source code for the YASS image steganography scheme. 

\bibliography{paper}
\bibliographystyle{abbrv}

\end{document}